\documentclass[pra,twocolumn,nofootinbib]{revtex4-2} 

\usepackage{amsmath}  
\usepackage{amsfonts} 
\usepackage{graphicx} 
\usepackage{hyperref}
\usepackage{amsthm}

\newcommand{\beq}{\begin{equation}}
	\newcommand{\eeq}{\end{equation}}
\newcommand{\bqa}{\begin{eqnarray}}
	\newcommand{\eqa}{\end{eqnarray}}
\newcommand{\nn}{\nonumber}

\newcommand{\smallfrac}[2]{\mbox{$\frac{#1}{#2}$}}

\newcommand{\half}{\smallfrac{1}{2}}

\begin{document}

\title{Why quantum correlations are shocking}

\author{Michael J. W. Hall}
\affiliation{Theoretical Physics, Research School of Physics, Australian National University, Canberra ACT 0200, Australia}



\begin{abstract}
A simple minimalist argument is given for why some correlations between quantum systems boggle our classical intuition.  The argument relies on two elementary physical assumptions, and recovers the standard experimentally--testable Bell inequality in a form that applies equally well to correlations between six--sided dice and between photon polarizations. The first assumption, that measurement selection in a first lab leaves the measurement statistics in a remote lab invariant (no--signaling), has been empirically verified, and is shown to be equivalent to the existence of a corresponding joint probability distribution for quantities measured in the first lab. 
The observed violation of the Bell inequality is then equivalent to the failure of a second assumption, that measurement selection in the remote lab leaves such a joint distribution invariant. Indeed, the degree of violation lower--bounds the variation of the joint distribution.  It directly follows there are just three possible physical mechanisms underlying such violations --- action--at--a--distance (superluminality), unavoidable common factors linking measurement choice and distant properties (conspiracy), and intrinsically incompatible physical quantities (complementarity). The argument extends to all Bell inequalities, and is briefly compared with other derivations.  
\end{abstract}

\maketitle 

\section{Introduction} 
\label{sec:intro}

Niels Bohr is famously quoted as saying that ``Those who are not shocked when they first come across quantum theory cannot possibly have understood it''~\cite{bohrshock}.
The nature of correlations between distant quantum systems is particularly shocking, as repeatedly demonstrated in a long line starting from the Einstein--Podolsky--Rosen argument for the incompleteness of quantum mechanics~\cite{epr}, passing through John Bell's demonstration that quantum mechanics is incompatible with a statistical inequality  for local realistic theories~\cite{bell1964}, to the experimental violation and remarkable applications of further such Bell inequalities~\cite{bellreview}, and most recently leading to the awarding of the 2022 Nobel Prize for Physics for ``experiments with entangled photons, establishing the violation of Bell inequalities and pioneering quantum information science''~\cite{nobel}.

There are several nice expositions on the remarkable nature of quantum correlations that derive forms of Bell's original 1964 inequality\cite{wigner,mermin1981} (or illuminate other aspects of such correlations~\cite{mermin1990,feynman1982,hardy2000,aravind2004}), which rely on an assumption of perfect correlations between measurement outcomes --- something that cannot be achieved in practice~\cite{bertlmann}.
In contrast, the well--known Clauser-Horne-Shimony-Holt (CHSH) Bell inequality~\cite{chsh}  does not require perfect correlations,
and its violation by quantum systems has been observed in many experiments~\cite{clauser1972,aspect1982,wineland2001,gisin2004,loop1,loop2,loop3,storz2023}, including photon polarization experiments suitable for advanced undergraduates~\cite{dehlinger2002,negre2023}. Such real--world violations are not only of fundamental physical interest, but also provide an essential resource for performing classically impossible tasks in quantum cryptography and random number generation~\cite{bellreview}.

However, the simplest approaches in the literature for deriving the CHSH Bell inequality rely on a restrictive assumption of predetermined or counterfactual values for measurable quantities~\cite{eberhard,scarani,gill,dehlinger2002,footdeterm}.  Further, while the inequality may alternatively be derived from various general sets of  assumptions~\cite{chsh,bellreview,bertlmann,fine1}, including for example hidden variables, local causality, parameter independence, outcome independence, measurement independence, and/or the existence of formal joint probability distributions, the abstract nature of these assumptions leads to seemingly endless debates about the physical significance of each. Hence there is value in providing an elementary and direct derivation that does not rely on any of the above assumptions.

To emphasise that mere {\it formal} simplicity is not enough, suppose that four measurable quantities $A, A', B, B'$ have a joint probability distribution for their possible measurement outcomes. At most three of the propositions $A=B, A=B', A'=B$ and $A'\neq B'$ can hold for any member of an ensemble generated by this distribution (since the first three imply $A'=B=A=B'$, which contradicts the fourth),
i.e., the arithmetic sum of their truth values is no more than 3 for any member. Averaging this sum over the ensemble then gives 
\beq \label{chsh1}
p(A=B)+p(A=B')+p(A'=B) +p(A' \neq B') \leq 3 , 
\eeq
which is a form of the celebrated CHSH Bell inequality~\cite{bellreview,zuk2014} (see also Sec.~\ref{sec:bell}).  Yet although this derivation is very simple and general, it only allows one to conclude that if measured probabilities on the left hand side violate the inequality then there is no formal joint probability distribution for $A, A', B, B'$ --- the physical meaning of which is not immediately apparent, nor the physical contexts in which it might be expected to hold.

\begin{figure*}[!t]
	\centering
	
	\includegraphics[width=
	0.8\textwidth]{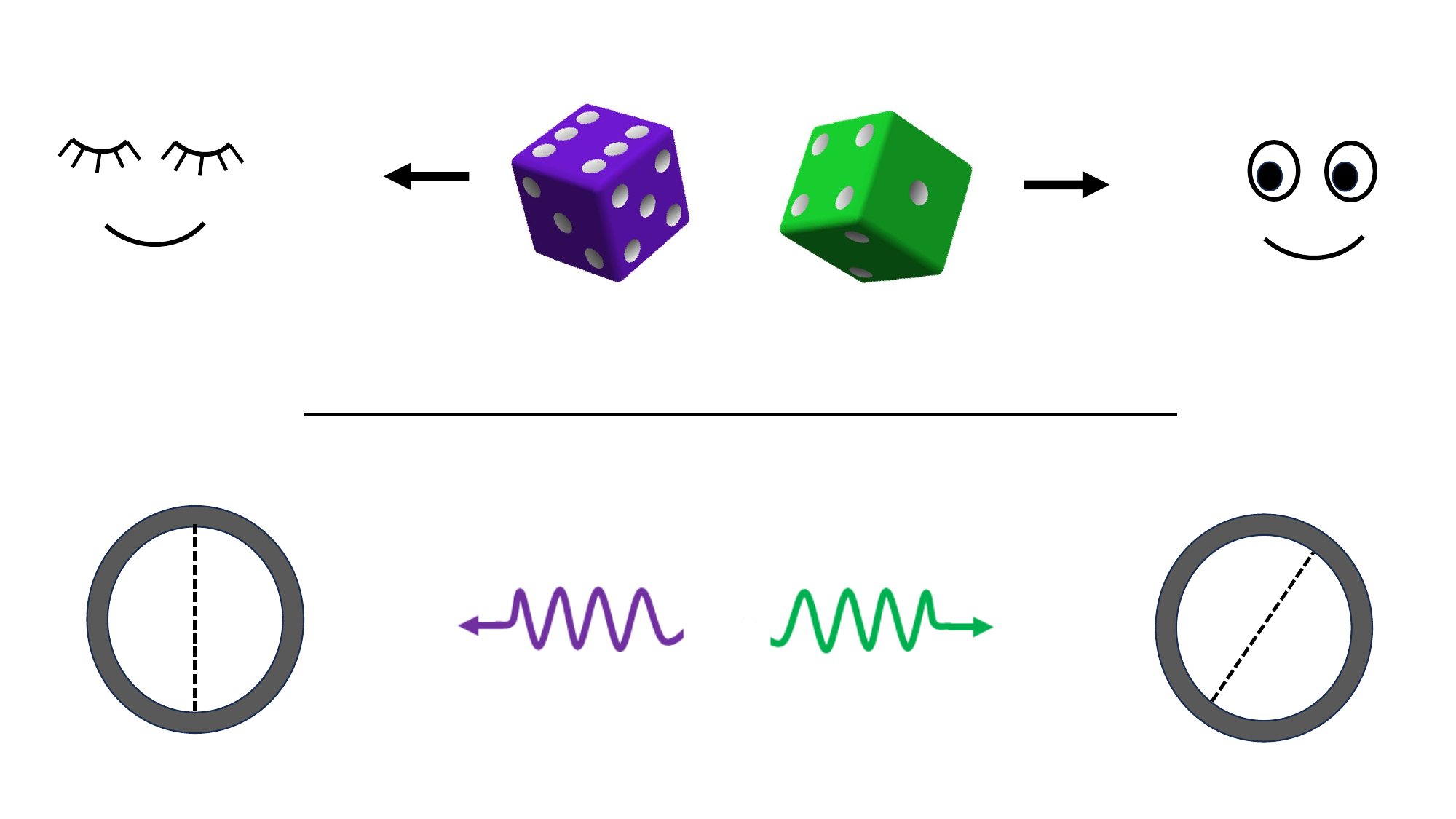}
	
	\caption{Two examples of joint experiments. In the upper panel Alice and Bob each receive a six-sided die on each run, roll their die with their eyes open or closed, and record the outcome. In the lower panel Alice and Bob each receive a photon on each run, orient a polarizer at one of two angles to the vertical, and record whether or not their photon passes through the polarizer. By making many runs they can estimate the probability that they each get the same outcome, for each of the four possible joint measurement settings (eyes open/closed, or polarizers oriented at first/second angles, on each side). These probabilities quantify the correlations between the outcomes, and in some cases have highly counterintuitive properties.
	}
	\label{fig1}
\end{figure*}

With the above in mind, the CHSH Bell inequality~(\ref{chsh1}) is derived here from two minimal and physical assumptions, building on a result of Fine~\cite{fine1}.
The approach is motivated in Sec.~\ref{sec:con} via a basic conundrum: if certain correlations are observed, then the probability that two possible measurements in one lab have equal outcomes appears to depend upon the choice of measurement in a distant lab. This conundrum  is sharpened in Sec.~\ref{sec:bell} to yield a simple yet rigorous derivation of the CHSH Bell inequality, based on two very intuitive assumptions related to the invariance of probabilities in one lab with respect to actions carried out in the other lab, where one of these assumptions (no--signaling) has been empirically verified. Observed violations of the inequality then correspond to violations of the second assumption (joint invariance), implying our world is highly nonintuitive: its explanation requires one of  faster--than--light influences, common factors that correlate measurement selection with remote properties, or intrinsically incompatible measurable quantities, as discussed in Sec.~\ref{sec:imp}.  
Brief concluding remarks are made in Sec.~\ref{sec:disc}, and the generalization of  the derivation and its close connections  with Fine's general approach to Bell inequalities~\cite{fine1,fine2} are noted in an Appendix.

\section{The basic conundrum}
\label{sec:con}

Suppose that two experimenters, Alice and Bob, perform a joint experiment of the following type.  Alice measures one of two measurable quantities or observables, $A$ and $A'$, on each run of the experiment. For example, Alice may receive a six-sided die on each run, with $A$ referring to the outcome of rolling the die with her eyes open and $A'$ to the outcome of rolling the die with her eyes closed, with possible outcomes 1, 2, \dots, 6. In a quantum example Alice may receive a photon on each run, with $A$ referring to whether it passes through a polarizer that is oriented at some angle  to the vertical and $A'$ to  whether it passes the polarizer when oriented at some different angle, with possible outcomes $y$ or $n$.

Similarly, Bob, located in a distant lab, measures one of two measurable quantities, $B$ and $B'$, on each run of the experiment. For example, Bob may receive and roll a second die on each run, with his eyes open or closed, or measure whether a second photon passes through a polarizer oriented at one of two different angles (see Fig.~\ref{fig1}).

Experiments of the above type are sufficient for motivating why some correlations can surprise or even shock our intuition. As an extreme example, suppose that whenever Alice measures the quantity $A$ then Bob obtains the same outcome as Alice, whether he measures  $B$ or $B'$. It thus seems reasonable to conclude that  $B=B'$ {\it when Alice measures} $A$. Suppose further, however, that whenever Alice measures the quantity $A'$ then Bob obtains the same outcome as Alice if he measures $B$ but a different outcome  if he measures $B'$. It is then just as reasonable to conclude that  $B\neq B'$ {\it when Alice measures} $A'$.  But how can the truth or otherwise of $B=B'$ in Bob's lab be dependent on Alice's choice of measurement in a distant lab, e.g., on whether she opens or closes her eyes? Clearly something {\it un}reasonable, or at least unexpected, is going on!~\cite{prbox}

Weaker correlations --- including quantum correlations --- can be just as unexpected. To see this, assume that Alice and Bob get together after many runs of an experiment of the above type, and calculate some measurement statistics. 
In particular, they estimate the probability $p(A=B)$ of obtaining equal outcomes, from the subset of runs where Alice measured $A$ and Bob measured $B$, and they similarly estimate the probabilities $p(A=B')$, $p(A'=B)$ and $p(A'\neq B')$ from the corresponding subsets of runs. We then have a basic conundrum that lies at the heart of the CHSH Bell inequality: {if the probabilities $p(A=B)$ and $p(A=B')$ are sufficiently high then $p(B=B')$ is expected to be high, while if $p(A'=B)$ and $p(A'\neq B')$ are also sufficiently high then $p(B\neq B')$ is expected to be high}. 

A physically testable example of this conundrum is provided by the photon polarization scenario depicted in Fig.~\ref{fig1}.  In particular, experimental data  shows that it is possible to physically prepare the photons on each run such that the measured probabilities $p(A=B)$, $p(A=B')$, $p(A'=B)$ and $p(A'\neq B')$ are each greater than $75\%$~\cite{aspect1985,negrefoot,footother}.  
But if the probabilities of two propositions  each occupy more than 75\% of a sample space, then their common overlap must occupy over 50\% of the sample space (since the first proposition leaves less than 25\% of the sample space unoccupied that might be occupied by the second). Noting that $B=B'$ in the common overlap of $A=B$ and $A=B'$, it follows that
\bqa \label{75a}
&p(A=B)> 75\% ~~{\rm and}~~p(A=B')> 75\% \nn\\  &{\rm implies}~~p(B=B')> 50\% 
\eqa
for a measurement of $A$. Similarly, $B\neq B'$ in the common overlap of $A'=B$ and $A'\neq B'$, yielding
\bqa \label{75b}
&p(A'=B)> 75\% ~~{\rm and}~~p(A'\neq B')> 75\% \nn\\  &{\rm implies}~~p(B\neq B')> 50\%  
\eqa
for a measurement of  $A'$. Thus, to avoid the contradiction $p(B=B')+p(B\neq B')>1$ in these real--world experiments, it seems that the probability of $B=B'$ is influenced by Alice's choice of polarizer angle in a distant lab! As will be seen, however, there is a hidden assumption in the above reasoning,  and alternative but equally surprising conclusions can be made.

The next section clarifies and strengthens the basic conundrum  to obtain the CHSH Bell  inequality~(\ref{chsh1}). Further,  the assumptions underpinning the derivation are clearly identified, leading to a discussion in Sec.~\ref{sec:imp} of the possible physical mechanisms that permit quantum correlations to shock our classical intuition.

\section{Minimalist derivation of the \\
CHSH Bell inequality}
\label{sec:bell}

The derivation of the Bell CHSH inequality~(\ref{chsh1}) given here relies on two  natural minimal assumptions (connected to those used in Proposition~1 of Fine~\cite{fine1}). The first is that Bob cannot signal to Alice via his choice of measurement. As will be discussed, this ``no-signaling'' assumption has the virtues of being experimentally verified and of being equivalent to the existence of joint probability distributions for Bob's observables that are compatible with the measured correlations, leading to a simple yet rigorous generalization of the conundrum posed by Eqs.~(\ref{75a}) and~(\ref{75b}).  The second assumption has a similar flavour, requiring that there is at least one such joint distribution that is invariant with respect to Alice's choice of measurement. It leads directly to the CHSH Bell inequality, implying that the experimental violation of this  inequality is due to the failure of this ``joint invariance'' assumption. 
The derivation throws significant light on the physical consequences of violating Bell inequalities (discussed in Sec.~\ref{sec:imp}).

\subsection{The no--signaling assumption and \\joint compatibility}
\label{ass1}

The basic conundrum of the previous section, exemplified in Eqs.~(\ref{75a}) and~(\ref{75b}), refers to the probabilities of $B=B'$ and $B\neq B'$, and hence implicitly requires that such probabilities  exist. This existence is nontrivial for the joint experiments depicted in Fig.~\ref{fig1}, since $B$ and $B'$ are not jointly measured and indeed are physically incompatible in these experiments. In particular,  Bob cannot roll his die with both eyes simultaneously open and shut, nor can he simultaneously orient his polarizer in two different directions. Remarkably, however, the existence of the required joint probabilities for $B$ and $B'$ is assured by the following physical and testable assumption.
{\flushleft \bf Assumption 1: No signaling from Bob to Alice.} {\it Alice's measurement statistics are invariant with respect to Bob's measurement selection.}
~\\

This no--signaling assumption is well--motivated whenever the two labs are sufficiently separated, and is consistent with both quantum and relativistic predictions~\cite{bellreview}. Importantly, it has been experimentally verified for quantum correlations~\cite{footnosig},  by checking that the measured joint distributions $p(a,b), p(a,b'), p(a',b), p(a',b')$  satisfy
\beq \label{nosig}
\begin{array}{c}
	p(a):= \sum_{b} p(a,b) =\sum_{b'} p(a,b'), \\~\\ p(a'):=\sum_{b} p(a',b)= \sum_{b'} p(a',b') ,
\end{array}
\eeq
up to statistical errors, where $a,a',b,b'$ denote outcomes of $A,A',B,B'$ respectively. The first  of these equations states that Alice's statistics for $A$  do not depend on whether Bob measures $B$ or $B'$, and the second states the same for $A'$. Thus Bob cannot signal to Alice by, e.g.,  measuring $B$ for many runs to transmit a ``0'', or measuring $B'$ to transmit a ``1'': Alice's statistics 
will be the same regardless of which measurement Bob performs.

The existence of the required joint probabilities for the conundrum, even for physically incompatible $B$ and $B'$, follows from the equivalence of no--signaling from Bob to Alice to the following property.
{\flushleft \bf Joint compatibility property:}  {\it The measurable quantities $A$, $B$ and $B'$ have a joint probability distribution compatible with the measured joint probability distributions, as do $A'$, $B$ and $B'$.}  
~\\

Here, denoting these joint distributions by $q(a,b,b')$ and $q'(a',b,b')$,\footnote{The labels $q$ and $q'$ correspond to Alice measuring $A$ and $A'$, respectively, where these labels are used to clearly distinguish $q(a,b,b')$ and $q'(a',b,b')$ from the measured distributions $p(a,b), p(a,b'), p(a',b), p(a,b')$.} ``{compatible with}'' means that the measured distributions $p(a,b), p(a,b'), p(a',b), p(a',b')$ are recovered as marginal distributions of $q(a,b,b')$ and $q'(a',b,b')$, i.e., that
\beq \label{marg}
\begin{array}{l}
	p(a,b)=q(a,b)=\sum_{b'} q(a,b,b'), \\~\\ p(a,b')=q(a,b')=\sum_{b} q(a,b,b'),\\~\\
	p(a',b)=q'(a',b)=\sum_{b'} q'(a',b,b'),\\~\\  p(a',b')=q'(a',b')=\sum_{b} q'(a',b,b').
\end{array}
\eeq 

To show the equivalence of the no--signaling assumption and the joint compatibility property, note first that the latter implies, via Eq.~(\ref{marg}), that
\begin{align*} \label{nosigba}  \nn
 \sum_b p(a,b) &= \sum_{b,b'} q(a,b,b') = \sum_{b'} p(a,b') ,\\
\sum_b p(a',b) &= \sum_{b,b'} q'(a',b,b')  = \sum_{b'} p(a',b') ,
\end{align*}
and hence no-signaling conditions~(\ref{nosig}) are satisfied.   Conversely, if the  no-signaling  conditions~(\ref{nosig}) are satisfied then, for example,  the joint probability distributions
\beq \label{existba}
\tilde q(a,b,b'):= \frac{p(a,b)p(a,b')}{ p(a)},~ \tilde q'(a',b,b'):= \frac{p(a',b)p(a',b')}{ p(a')} 
\eeq
are easily checked to satisfy the joint compatibility conditions~(\ref{marg}), as required (as do other examples~\cite{footq0}). The results derived below hold for any and all joint distributions $q$ and $q'$ satisfying joint compatibility, i.e., Eq.~(\ref{marg}) (with their existence being equivalent to the no--signaling assumption). The question of their physical significance is left until Sec.~\ref{sec:imp}. 

\subsection{Generalization of the basic conundrum via the no--signaling assumption}

The basic conundrum in Sec.~\ref{sec:con} can now be generalized and made rigorous, as a consequence of the no--signaling assumption, based on 
elementary properties of joint probabilities.
In particular, similarly to the derivation of Eq.~(\ref{chsh1}), note for any member of an ensemble generated from a joint probability distribution $q(a,b,b')$ (guaranteed to exist by the no--signaling assumption), that at most two of $A=B$, $A=B'$ and $B\neq B'$ can hold. Thus the arithmetic sum of their truth values is at most 2, and averaging this sum over the ensemble  gives  $q(A=B)+q(A=B')+q(B\neq B')\leq 2$, implying immediately from Eq.~(\ref{marg}) that
\beq \label{simp1}
q(B=B') \geq p(A=B)+p(A=B')-1,
\eeq 
generalizing Eq.~(\ref{75a}).  Similarly, replacing $A,B,B'$ by $B,A',B'$ and $q$ by $q'$ in the argument yields  
\beq \label{simp2}
q'(B\neq B') \geq  
p(A'=B)+p(A'\neq B') -1 ,
\eeq
generalizing Eq.~(\ref{75b}). Equations~(\ref{simp1}) and~(\ref{simp2})  are central to deriving the CHSH Bell inequality, and may be obtained in various alternative ways~\cite{footaltern}.

Inequalities~(\ref{simp1}) and~(\ref{simp2}) immediately recover the basic conundrum of the previous section when the measured probabilities  $p(A=B), p(A=B'), p(A'=B)$ and $p(A'\neq B')$ are each greater than 75\%: $q(B=B')$ and $q'(B\neq B')$ are then both greater than 50\% in this case, i.e., equality of $B$ and $B'$ is likely if Alice measures $A$, whereas inequality is likely if Alice measures $A'$.
More generally,  adding Eqs.~(\ref{simp1}) and~(\ref{simp2}) and using $q'(B\neq B')=1-q'(B=B')$ gives
\begin{align}
q(B=B') - q'(B=B')  &\geq p(A=B)+p(A=B') \nn\\ 
&+  p(A'=B)+p(A'\neq B') - 3  \label{gencon}
\end{align}
for any and all $q$ and $q'$. Hence, one has a generalized conundrum whenever the sum of the experimental probabilities on the right hand side is greater than 3: one must then have $q(B=B')\neq q'(B= B')$ for any choice of $q$ and $q'$, i.e., the probability of equality of Bob's observables is correlated with the measurement selection made in a distant lab!

\subsection{The joint invariance assumption and the \\CHSH Bell inequality}

The generalized conundrum in Eq.~(\ref{gencon}) can be reformulated in terms of the following assumption.
{\flushleft \bf Assumption 2: Joint invariance for Bob.} {\it There is a joint probability distribution for Bob's measurements, compatible with the measured correlations, that is invariant with respect to Alice's measurement selection.}
~\\

Thus, there are joint probability distributions $q(a,b,b')$ and $q'(a',b,b')$ (as guaranteed by the equivalence of joint compatibility and no--signaling),  such that the joint distributions $q(b,b')=\sum_a q(a,b,b')$ and $q'(b,b')=\sum_{a'}q'(a',b,b')$ of Bob's observables satisfy
\beq \label{indirect}
q(b,b') = q'(b,b') .
\eeq
Note that summing this condition over $b'$ ($b$) immediately implies that Bob's measurement statistics for $B$ ($B'$) are invariant with respect to Alice's  measurement selection, i.e., that there is no--signaling from Alice to Bob. However, joint invariance is a stronger requirement, since it also constrains the {\it joint} distribution of $B$ and $B'$. 
It nevertheless has a similar flavour to no--signaling, and likewise appears natural if the separation between Alice and Bob is sufficiently large
(or if Alice is free to choose her measurement independently of any physical factor that might influence Bob's joint distribution). 
Its physical significance is discussed further in Sec.~\ref{sec:imp}.


Substituting joint invariance condition~(\ref{indirect}) into Eq.~(\ref{gencon})  immediately yields the form of the CHSH Bell inequality in Eq.~(\ref{chsh1}):
\beq \label{bellprob}
p(A=B)+p(A=B')+p(A'=B) +p(A' \neq B') \leq 3 . 
\eeq
Note that it is valid for any number of outcomes for each measurable quantity, and whether or not the ranges of these outcomes overlap~\cite{footfine}. It is more typically written in the form~\cite{bellreview}
\begin{align} \label{chsh}
S_{AA'BB'} &:=E(A,B) + E(A,B') + E(A',B) - E(A',B')\nn\\
&~ \leq 2 ,
\end{align}
where $E(A,B)$ is the measure of correlation  defined by
\beq \label{eab}
E(A,B):= p(A=B) - p(A\neq B) .
\eeq
Thus $E(A,B)=1$ when $A$ and $B$ are always equal (perfect correlation), and $E(A,B)=-1$ when they are never equal (perfect anticorrelation). The equivalence of Eqs.~(\ref{bellprob}) and~(\ref{chsh}) follows immediately from the linear relation $p(A=B)=\half[1+E(A,B)]$. Further inequalities, such as $S_{A'ABB'}\leq 2$, may be obtained via permutations of quantities and outcomes~\cite{footperm}.

The CHSH Bell inequality~(\ref{bellprob}) is clearly violated if each of the probabilities is greater than 75\%, as expected from Eqs.~(\ref{75a}) and~(\ref{75b}), but it can also be violated more generally (e.g., for measured probabilities 80\%, 80\%, 80\%, and 70\%). The maximum in--principle violation corresponds to each of the four measured probabilities being equal to 1, as per the extreme example discussed in Sec.~\ref{sec:con}. However, as is well known, for quantum systems the maximum possible violation is smaller, corresponding to each of  the four measured probabilities being equal to $\half(1+1/\sqrt{2})\approx 85.36\%$, yielding  maximum values of $2+\sqrt{2}\approx 3.4142$ and $2\sqrt{2}\approx 2.8284$ for the left hand sides of inequalities~(\ref{bellprob}) and~(\ref{chsh}) respectively~\cite{bellreview}.

Crucially, violations of the CHSH Bell inequality have been observed in several loophole--free experiments~\cite{loop1,loop2,loop3,storz2023}, which have very high detection efficiencies (avoiding the need to further assume that the statistics are not skewed by a detection bias), and with the labs located at a sufficient distance to prevent any sub--lightspeed influences propagating between them on each run (avoiding the need to further assume the statistics are not skewed by such influences). Recalling that such experiments also show that the no--signaling assumption is valid~\cite{footnosig}, this immediately implies that the joint invariance assumption is physically invalid. This has significant consequences, as discussed in the next section.

\section{Physical implications}
\label{sec:imp}

As noted above, loophole--free experimental violations of the CHSH Bell inequalities~(\ref{bellprob}) and~(\ref{chsh}) imply that the joint invariance assumption fails: any joint probability distribution of Bob's observables compatible with the measurement statistics (guaranteed to exist via the equivalence of no--signaling and joint compatibility), is correlated with the selection of measurement in Alice's lab.  This has a number of interesting implications, some of which are briefly discussed here. 

First, the violation of the CHSH Bell inequality by a given set of correlations immediately implies there is no local deterministic or local causal model for the correlations, in which the statistics of $A,A',B,B'$ are determined by local probabilities $p(a|\lambda), p(a'|\lambda), p(b|\lambda), p(b'|\lambda)$ that depend on the value of a `hidden' variable $\lambda$ having some prior probability distribution $p(\lambda)$~\cite{bell1964,bellreview,chsh,bertlmann}. In particular, for any such model one may take
\beq \label{ld}
\begin{array}{l}
	q_{\rm L}(a,b,b') = \sum_\lambda p(\lambda) 
	p(a|\lambda) p(b|\lambda) p(b'|\lambda) ,
	\\~\\ 
	q'_{\rm L}(a',b,b') = \sum_\lambda p(\lambda)
	p(a'|\lambda) p(b|\lambda) p(b'|\lambda)  
\end{array}
\eeq
(replacing sums by integrals for continuous ranges), for which $q_{\rm L}(b,b')=q'_{\rm L}(b,b')=\sum_\lambda p(\lambda)  p(b|\lambda) p(b'|\lambda)$. Thus joint invariance condition~(\ref{indirect}) holds, implying the CHSH Bell inequality holds for any such model of the correlations. It is similarly straightforward to show that quantum correlations can violate the inequality only if Bob's observables are not compatible~\cite{footincompat}.

Second, there is a simple numerical link between the degree of violation of the CHSH Bell inequality and the degree of violation of joint invariance condition~(\ref{indirect}). In particular, the generalized conundrum in Eq.~(\ref{gencon}) is equivalent to
\beq  \label{lowerb}
	q(B=B') - q'(B=B')  \geq  \half(S_{AA'BB'}-2) ,
\eeq
where $S_{AA'BB'}$ denotes the left hand side of CHSH Bell inequality~(\ref{chsh}). Thus, the amount of violation sets a lower bound on the variation of the probability of $B=B'$ with Alice's measurement selection.   For example, for the maximum quantum violation  $S_{AA'BB'}=2\sqrt{2}$ one has $q(B=B') - q'(B=B')\geq \sqrt{2}-1\approx 41\%$ for all $q$ and $q'$. It may be shown that the quantity  $\half(S_{AA'BB'}-2)$ is also a lower bound for the separation between the joint distributions $q(b,b')$ and $q'(b,b')$~\cite{footaltbell}.

Third, noting the critical role of the joint invariance assumption, it is natural to consider how the experimental failure of this assumption, implied by the violation of the CHSH Bell inequality, can be interpreted. There are essentially just three possibilities in this regard, examined in turn below.  

The first possibility for explaining the failure of joint invariance in loophole--free experiments is that Alice's choice of measurement transmits a faster--than--light physical influence that acts on the joint statistics of Bob's observables (``superluminality''). A simple model of this sort, for the photon polarization example in Fig.~\ref{fig1}, has been given by Toner and Bacon~\cite{toner}. Such influences are, however, difficult to reconcile with notions of relativistic causality. 

The second possibility is that the unavoidable correlation between measurement choice in one lab and joint probabilities in a distant lab is due to a common cause (or retrocause) ---  some shared physical factor or variable that acts to correlate Alice's measurement selection with the joint statistics of Bob's observables (``conspiracy'').  Simple models of this sort are known for the photon polarization example in Fig.~\ref{fig1}~\cite{shimony,hallfree}, and such a model may also be given for a maximal violation of the CHSH Bell inequality in the context of the dice-rolling example in Fig.~\ref{fig1}, with the dice colors providing the common physical factor~\cite{footfree}. This possibility has been described by Bell as ``even more mind boggling than one in which causal chains go faster than light''~\cite{bertlmann}, but it nevertheless merits equally serious consideration~\cite{palmer,bigbelltest}.

The third (and rather different) possibility is that the joint invariance assumption concerns unphysical distributions, so that its failure is only a formal issue (``complementarity''). In particular, the existence of joint distributions $q(b,b')$ and $q'(b,b')$ is guaranteed by the no--signaling assumption, but these distributions can be regarded as purely formal if there is no method of directly estimating them as the relative frequencies of some joint measurement, i.e., if $B$ and $B'$ are intrinsically incompatible.  This possibility is consistent with standard quantum mechanics, since Bell inequalities can only be violated by quantum systems if there is no accurate joint measurement of Bob's observables within the theory~\cite{fine2,footincompat}. 

Finally, the astute reader may have noted the standard rules of probability have been assumed to apply throughout. It is therefore possible to model Bell inequality violation by allowing the joint distribution of $B$ and $B'$ to be invariant but nonclassical in some way, e.g., by allowing it to take negative values~\cite{feynman1982,neg,higgins}. Note, however, that this possibility is just an alternative expression of complementarity, as it again rules out direct estimates of the joint distribution as relative frequencies of some joint measurement~\cite{indirect}.  Instead, such negative values act as a useful marker of intrinsic incompatibility, with a number of related applications~\cite{quasi}.

The above three possibilities arise solely from the form and failure of the joint invariance assumption, without any reference to concepts or assumptions typically considered in hidden--variable approaches to Bell inequality violation (indeed there is not even a hidden--variable decomposition to base a comparison on). This has the advantage of avoiding the (sometimes contentious) interpretations of such concepts. However, it may be noted that the three options of superluminality, conspiracy and complementarity broadly capture notions expressed by the respective failures of parameter independence, measurement independence and determinism in hidden--variable approaches~\cite{bellreview}.
More general discussion on the implications of Bell inequality violation, including for the interpretation of quantum mechanics, may be found elsewhere~\cite{laloe,stanford}.

\section{Conclusions}
\label{sec:disc}


The nature of quantum correlations has been called ``mind boggling''~\cite{hardy2000,bertlmann}, and the main aim of this paper has been to make the underlying reasons transparent  via an approach that uses minimal assumptions. The basic conundrum described in Sec.~\ref{sec:con} makes clear what type of correlations shock our classical intuition; the sharpening and generalization of this conundrum in Sec.~\ref{sec:bell} then leads to a simple derivation of CHSH Bell inequality~(\ref{chsh}), via two assumptions related to  the invariance of probabilities in one lab with respect to the measurements made in a distant lab. No other assumptions (e.g., counterfactuality, hidden variables, local causality, measurement independence, etc.) are required, making the physical implications of the violation of the inequality straightforward, as discussed in Sec.~\ref{sec:imp}. The generalization to arbitrary numbers of measurements on each side and the close connection of the assumptions with Fine's approach to Bell inequalities are discussed in Appendix~\ref{sec:app}.

The first assumption, that  Alice's measurement statistics are invariant with respect to Bob's measurement selection (no--signaling), has been verified by experiment and implies that joint probability distributions exist for Bob's measured quantities (even when they are physically incompatible). Hence the observed violations of the CHSH Bell inequality imply it is the second assumption, that there is such a joint probability distribution which is invariant with respect to Alice's measurement selection (joint invariance), that fails. Further, the degree of violation of the inequality directly determines the degree to which Bob's joint probability distribution must vary with Alice's measurement selection, as per Eq.~(\ref{lowerb}).  
 
A virtue of the CHSH Bell inequality is that it requires no assumptions about the validity of quantum theory: the latter is only relevant in guiding the design of corresponding experiments. Hence, the derivation in Sec.~\ref{sec:bell}, coupled with  experimental results, implies that the joint invariance assumption must be rejected in {\it any} physical theory of the observed correlations, making the implications discussed in Sec.~\ref{sec:imp} applicable to any such theory. 

It is natural to ask, following the discussion in Sec.~\ref{sec:imp}, which is the least implausible option available for explaining Bell inequality violations: faster--than--light influences (superluminality), common causes correlating measurement choices with system properties (conspiracy), or the intrinsic incompatibility of some measurable quantities (complementarity)? Each one is shocking to a classical palate, making the answer largely a matter of taste.  Nevertheless, the observed violations --- even seen, remarkably enough, in undergraduate--level experiments~\cite{dehlinger2002,negre2023} --- implies that one such possibility must be swallowed!
 
In this regard it is probably fair to say that most physicists working in quantum information theory (and most adherents of the completeness of quantum mechanics) would prefer the complementarity option. In particular, unlike the superluminal and conspiratorial options, any joint probability distribution of $B$ and $B'$ is thereby regarded as unphysical, not corresponding to the statistics of any accurate joint measurement of these quantities, thus making the variation of such distributions with Alice's measurement selection physically moot (see Sec.~\ref{sec:imp}). This has the advantages of being able to consistently maintain the speed of light as an upper bound on the propagation of physical influences; to permit experimental choices to be independent of system properties; and to allow Bell inequality violation to be exploited  as a resource for classically--impossible tasks. In particular, under this option such violations preclude an external eavesdropper or other physical system from gaining complete information about outcomes via an accurate joint measurement, allowing Alice and Bob to certifiably generate secure cryptographic keys and truly random numbers~\cite{bellreview}.

Finally, the approach here suggests several future research directions. For example, no--signaling from Bob to Alice as per the first assumption is again natural (and experimentally verified) when Alice's measurements are made in the past of Bob's measurements. Hence the latter have a joint distribution, and Leggett-Garg  and temporal Bell inequalities~\cite{lgreview} can then be obtained by replacing the usual macrorealism and noninvasive measurability assumptions with joint invariance (reinterpreted as a nondisturbance assumption). The approach may also be extendable to noncontextuality inequalites on hypergraphs~\cite{noncon}, with the no-signaling and joint invariance assumptions reinterpreted in terms of nondisturbance. It would further be of interest to investigate maximally efficient models of CHSH Bell inequality violation, in the sense of saturating the bound in Eq.~(\ref{lowerb}).

\acknowledgements I am grateful to Y. Chua, S. Gananadha and C. Lim for 
critical resources, and to Arthur Fine for several excellent suggestions.

\appendix
\section{Generalization and connection with Fine's general approach to Bell inequalities}
\label{sec:app}

The approach in the main text is easily extended to general Bell inequalities, based on its connections to Fine's characterization of such inequalities in terms of joint probability distributions~\cite{fine1,fine2}.

Proposition~1 of Ref.~\cite{fine1} considers the case of two measurement choices in each lab, while Theorem~1 of Ref.~\cite{fine2} generalizes to arbitrary numbers of measurement choices. 
In the latter case, suppose on each run of a joint experiment that Alice measures one of $m$ measurable quantities $A_1,A_2,\dots,A_m$ and Bob measures one of $n$ measurable quantities $B_1,B_2,\dots,B_n$. After many runs they can then estimate the joint probability distribution $p(a_j,b_k)$ of outcomes $A_j=a_j$ and $B_k=b_k$, for each $j=1,2,\dots,m$ and $k=1,2,\dots,n$. 
The joint compatibility property in Sec.~\ref{ass1} then generalizes naturally as follows.
{\flushleft \bf Generalized joint compatibility property:} {\it The measurable quantities $A_j$, $B_1,\dots B_n$ have a  joint probability distribution $q_j(a_j,b_1,\dots,b_n)$ compatible with the measured probabilities, for each $j=1,\dots m$.}\\~\\
Fine's Theorem~1 states that the combination of generalized joint compatibility and joint invariance for Bob are equivalent to the existence of a formal joint probability distribution for $A_1,\dots,A_m, B_1,\dots B_n$ that is consistent with the measured probabilities~\cite{fine2}.  His proof is straightforward, but a modified version will be given here noting that, similarly to the main text, generalized joint compatibility is equivalent to the physically more transparent (and experimentally verified) equivalent assumption of no--signaling from Bob to Alice, i.e. to
\beq \label{nosiggen}
p(a_j) = \sum_{b_k} p(a_j,b_k)
\eeq
for each $j=1,2,\dots,m$ and $k=1,2,\dots,n$. One has, e.g., the example distributions
\beq \label{pj}
\tilde q_j(a_j,b_1,\dots,b_n) := \frac{p(a_j,b_1) \dots p(a_j,b_n)}{p(a_j)^{n-1}} ,
\eeq
in analogy to the example in Eq.~(\ref{existba}). Together with the joint invariance assumption that the marginal joint distribution $q(b_1,\dots b_n)$ of Bob's quantities  is invariant with respect to Alice's measurement choice $A_j$, i.e., that $q(b_1,\dots b_n)=\sum_{a_j} q_j(a_j,b_1,\dots,b_n)$ for each $j=1,\dots,m$, it follows that
\beq \label{pgen}
\small{q(a_1,\dots,a_m,b_1,\dots b_n) := \frac{q_1(a_1,b_1,\dots,b_n)\dots q_m(a_m,b_1,\dots,b_n)}{q(b_1,\dots,b_n)^{m-1}} }
\eeq
is a formally well--defined joint probability distribution for $A_1,\dots,A_m, B_1,\dots B_n$ that is compatible with the measured probabilities.  Conversely, given such a joint probability distribution then it is straightforward to check that no--signaling from Bob to Alice and joint invariance for Bob both hold.

General Bell inequalities are now defined as the set of nontrivial constraints imposed on the measured probabilities by the formal existence of such a classical joint probability distribution $q(a_1,\dots,a_m,b_1,\dots b_n)$~\cite{bellreview,fine1,fine2}. Any violation of such inequalities then immediately implies, recalling that the no-signaling assumption is experimentally verified, that the joint invariance assumption fails, leading to similar physical implications as discussed in Sec.~\ref{sec:imp} for the CHSH Bell inequality. 

Note that the CHSH Bell inequality~(\ref{bellprob}) may alternatively be obtained via the construction of a formal joint distribution $q(a,a',b,b')$ from the two assumptions, as per Eq.~(\ref{pgen}), by applying the method given in Sec.~\ref{sec:intro}. However, the derivation in Sec.~\ref{sec:bell} via Eqs.~(\ref{simp1}) and~(\ref{simp2}) has the advantage of being able to directly link the degree of violation of the inequality with the degree to which the joint invariance assumption fails, as per Eqs.~(\ref{gencon}) and~(\ref{lowerb}).


\begin{thebibliography}{99}
\bibitem{prbox} If $A, A',B$ and $B'$ each have two possible outcomes, both occurring with equal probability $\half$, this example corresponds to the well--studied case of  ``PR--box'' correlations~\cite{bellreview,rastall,pr}. Note that local deterministic models of such extreme correlations exist (an example will be given in the context of the dice-rolling experiment, in which the outcome statistics are correlated with measurement choices via the colors of the dice).
\bibitem{footq0}   Other suitable joint distributions $q$ and $q'$ include, e.g., $q_C(a,b,b')=\sum_c p(a,b,c) p(a,b',c)/p(a,c)$ and  $q'_{C'}(a',b,b')=\sum_{c'} p(a',b,c') p(a',b',c')/p(a',c')$, where $C$ and $C'$ are any variables comeasurable with $A$ and $A'$ respectively such that $p(a,c)$ and $p(a',c')$ are invariant with respect to Bob's measurement selection. A further example, for local causal models, is given in Eq.~(\ref{ld}).
\bibitem{footaltern} It may be helpful to readers to give two alternative derivations of fundamental inequalities~(\ref{simp1}) and~(\ref{simp2}). For example, note that $B\neq B'$ implies that $A\neq B$ or $A\neq B'$, and hence one has  $q(B\neq B')\leq q(A\neq B~{\rm or}~A\neq B')\leq q(A\neq B)+q(A\neq B')$. This ``triangle inequality'' for the measure of distance $d(A,B):=q(A\neq B)$ is equivalent to inequality~(\ref{simp1}), and one similarly has  $q'(A'\neq B')\leq q'(A'\neq B)+q'(B\neq B')$ which is equivalent to inequality~(\ref{simp2}). As a second alternative derivation, analogous to that of Eqs.~(\ref{75a}) and~(\ref{75b}), note that in any joint sample space of $A,B,B'$ with distribution $q(a,b,b')$ (which exists via the no signaling assumption) there is only a fraction $1-q(A=B)$ available for $A=B'$ to occupy that is not already occupied by $A=B$, and hence their common overlap must occupy a fraction of at least $q(A=B')-[1-q(A=B)]=q(A=B)+q(A=B')-1$ of the sample space. But $B=B'$ in this common overlap,  implying Eq.~(\ref{simp1}) via Eq.~(\ref{marg}). Equation~(\ref{simp2}) similarly follows by considering the common overlap of $A'=B$ and $A'\neq B'$ for any joint distribution $q'(a',b,b')$.
\bibitem{footfine} For the case of two--valued measurable quantities with the same range of outcomes one can obtain a lower bound for the sum in CHSH Bell inequality~(\ref{bellprob}) by a  similar argument (which I am indebted to Arthur Fine for communicating to me). First, for $B\neq B'$ one must have $A=B$ or $A=B'$ in this case, implying $q(B\neq B')\leq q(A=B)+q(A =B')$. Similarly, for $B=B'$ one must have $A'=B$ or $A'\neq B'$, implying $q'(B=B')\leq q'(A'=B)+ q'(A'\neq B')$. Adding these inequalities and using joint invariance and Eq.~(\ref{marg}) yields  $p(A=B)+p(A=B') + p(A'=B)+ p(A'\neq B')\geq1$. For a combined range of more than two outcomes, however, there is only a trivial lower bound of zero (e.g., take $A\equiv1, B\equiv 2, A'\equiv B'\equiv 3$). Further inequalities for the two-valued and other cases are given below~\cite{footperm}.
\bibitem{footperm} Interchanging the labels $A$ and $A'$ and/or $B$ and $B'$ for Alice's and Bob's quantities leaves the assumptions invariant, yielding three further CHSH Bell inequalities $S_{A'ABB'}, S_{AA'B'B}, S_{A'AB'B}\leq 2$ in addition to Eq.~(\ref{chsh}). Permutations of outcome labels yield additional inequalities. For example, for the case of two possible outcomes for each quantity, let $\bar{A}$ denote the same quantity as $A$ but with the outcome labels swapped. Hence $p(A=B)=p(\bar A\neq B)$, implying $E(\bar A,B)=-E(A,B)$ from Eq.~(\ref{eab}), giving the identity $S_{\bar AA'BB'}=-S_{AA'B'B}$ (i.e., negating the first subscript of $S$ is equivalent to swapping the last two subscripts and changing the sign). Substitution into Eq.~(\ref{chsh}) and the above three inequalities then yields the further four CHSH Bell inequalities $S_{AA'BB'},S_{A'ABB'}, S_{AA'B'B}, S_{A'AB'B}\geq -2$. These eight inequalities in total, for the case of the same two possible outcomes, are in fact necessary and sufficient for the joint invariance assumption to hold and for the existence of a joint probability distribution $p(a,a',b,b')$~\cite{fine1} (see also the Appendix).
\bibitem{footincompat} For quantum observables $A,A',B,B'$ described by positive--operator--valued measures (POVMs) $\{E_a\}$, $\{E'_{a'}\}$, $\{F_b\}$,  $\{F'_{b'}\}$, suppose that $B$ and $B'$ are compatible, i.e., have a joint POVM $\{F_{bb'}\}$ satisfying $F_b=\sum_{b'}F_{bb'}$ and $F'_{b'}=\sum_{b}F_{bb'}$. One can then take $q_{\rm Q}(a,b,b')=\langle E_a\otimes F_{bb'}\rangle$ and $q'_{\rm Q}(a',b,b')=\langle E'_{a'}\otimes F_{bb'}\rangle$, yielding $q_{\rm Q}(b,b')=q'_{\rm Q}(b,b')= \langle F_{bb'}\rangle$. Thus joint invariance is satisfied, recovering the known result that Bob's observables must be incompatible if they are to violate the CHSH Bell inequality~\cite{fine2}. The same conclusion is obtained for $A$ and $A'$ by interchanging the roles of Alice and Bob and noting that $S_{AA'BB'}=S_{BB'AA'}$.
\bibitem{footaltbell} One has $\sum_{b,b'}|q(b,b')-q'(b,b')|\geq \sum_{b}|q(b,b)-q'(b,b)|\geq |\sum_b [q(b,b)-q'(b,b)]|=|q(B=B')-q'(B=B')| \geq \half(S_{AA'BB'}-2)$.
Note that combining  Eq.~(\ref{lowerb}) with the Bell inequality $S_{A'ABB'}\leq 2$, obtained by interchanging the labels of $A$ and $A'$ in Eq.~(\ref{chsh}), yields the further lower bound $|q(B=B')-q'(B=B')|\geq\half\max\{S_{AA'BB'},S_{A'ABB'}\}-1$. The same bounds apply to joint probabilities of $A$ and $A'$, noting that $S_{AA'BB'}=S_{BB'AA'}$ when the roles of  Alice and Bob are swapped.
\bibitem{footfree} A causal model for the extreme example discussed at the beginning of Sec.~\ref{sec:con}, in the context of the dice-rolling experiment depicted in Fig.~\ref{fig1}, is obtained by using the colors of the dice as a common cause that correlates measurement selection with system properties.
In particular, suppose each die is either purple or green on each run. When both dice are purple, they are weighted so as to give fixed but different outcomes (with equal probabilities for each of the $6\times5=30$ pairs of such outcomes). Otherwise, the dice are weighted to give the same fixed outcome (with equal probabilities for each of the 6 pairs of such outcomes). Finally, Alice and Bob dislike purple and love green, and so each naturally rolls with their eyes shut when they receive a purple die and with their eyes open otherwise. Thus Alice and Bob always obtain the same outcome if she measures $A$ and/or he measures $B$, and always obtain  different outcomes if she measures $A'$ and he measures $B'$. 
This gives a local deterministic model of the correlations having no faster-than-light influences (with $q(a,b,b') = \tfrac{1}{6}\delta_{ab}\delta_{ab'}$ and $q'(a',b,b')=\tfrac{1}{30}\delta_{a'b}(1-\delta_{a'b'})$). By construction, $q(B=B')=1$ if Alice measures $A$ (i.e., if her die is green), whereas $q'(B=B')=0$ if she measures $A'$ (i.e., if her die is purple). Thus the joint statistics of $B$ and $B'$ are correlated with Alice's measurement choice, and maximally violate the joint invariance condition~(\ref{indirect}) (and saturate the lower bound in Eq.~(\ref{lowerb})).
\end{thebibliography}
\end{document}